\documentclass[twocolumn]{aastex7}

\usepackage{xcolor}

\begin{document}

\title{Stellar Variability and Distance Indicators in the Near-infrared in Nearby Galaxies. II. Pulsating Stars in the Carina Dwarf Spheroidal}

\author[0000-0001-8771-7554,gname=Chow-Choong, sname=Ngeow]{Chow-Choong Ngeow}
\affil{Graduate Institute of Astronomy, National Central University, 300 Jhongda Road, 32001 Jhongli, Taiwan}
\affil{Taiwan Astronomical Research Alliance (TARA)}
\email[show]{cngeow@astro.ncu.edu.tw}

\author[0000-0001-6147-3360,gname=Anupam, sname=Bhardwaj]{Anupam Bhardwaj}
\affil{Inter-University Center for Astronomy and Astrophysics (IUCAA), Post Bag 4, Ganeshkhind, Pune 411 007, India}
\email{anupam.bhardwajj@gmail.com}

\author[0009-0004-7728-4600,gname=Prashant, sname=Nishad]{Prashant Nishad}
\affil{Inter-University Center for Astronomy and Astrophysics (IUCAA), Post Bag 4, Ganeshkhind, Pune 411 007, India}
\email{prashant.nishad@iucaa.in}

\author[0000-0003-3679-2428,gname=Das, sname=Susmita]{Susmita Das}
\affil{Inter-University Center for Astronomy and Astrophysics (IUCAA), Post Bag 4, Ganeshkhind, Pune 411 007, India}
\email{susmita.das@iucaa.in}


\begin{abstract}

  We present homogeneous, near-infrared ($JHK_s$ bands) time-series observations of the classical Carina dwarf Spheroidal (dSph) galaxy to determine accurate and precise distances using the pulsating stars as standard candles. These observations cover two Carina dSph fields ($\sim10.8\arcmin\times10.8\arcmin$) obtained with the FourStar infrared camera mounted on the 6.5-m Magellan Telescope. We collected precise photometric measurements of 43 RR Lyrae, 11 anomalous Cepheids (ACep), and 102 dwarf Cepheids (DCep) in Carina dSph. Using RR Lyrae, we obtained a distance modulus of $20.079\pm0.028\mathrm{(statistical)}\pm0.045\mathrm{(systematic)}$~mag, or a distance to Carina of $103.7\pm1.3\mathrm{(statistical)}\pm2.2\mathrm{(systematic)}$~kpc. The literature calibrations based on SX Phoenicis or delta-Scuti stars were used to anchor the $JHK_s$ period-luminosity relations for DCep. This resulted in a distance modulus that is in excellent agreement with RR Lyrae based determination. Finally, the distance moduli estimates using the ACep were found to be systematically smaller than the RR Lyrae-based distance modulus, suggesting a metallicity dependence on the ACep period-luminosity relation. 
  
\end{abstract}

\section{Introduction}

Dwarf galaxies in the local Universe are important laboratories providing valuable constraints for the studies of galaxy formation and evolution \citep[for reviews, see][]{hodge1971,tolstoy2009,mcconnachie2012}, as well as for testing the robustness of the widely accepted $\Lambda$CDM (cosmological constant plus cold dark matter) models \citep[e.g., see][]{sales2022}. A majority of these dwarf galaxies, especially the classical dwarf Spheroidal (dSph), have been extensively studied via multi-band imaging observations. These studies have revealed important physical properties of dSph galaxies including, but not limited to, distance ($D$ in kpc, using isochrone fitting or features such as the horizontal branch), physical size (in terms of half-light radius or tidal radius), ellipticity, the $V$-band absolute magnitude $(M_V)$, stellar content and the total stellar mass, star formation history (SFH) as well as photometric metallicities. 

Distance to a dSph galaxy is a fundamental and crucial parameter to constrain other physical parameters such as $M_V$ and radius. Yet, past distance measurements to classical dSph in the literature, using various techniques, exhibit a considerable scatter (with more than 5\% variations); see the discussion in \citet{Nagarajann2022}, \citet{oakes2022}, and \citet{bhardwaj2024} for a few representative classical dSph. The reasons behind such scatters are multifold, including the data quality (early vs. modern observations), possible hidden systematics associated with different techniques, as well as inconsistent calibrations of stellar standard candles used for distance measurements.

As a result, we initiated the Stellar VAriability and Distance Indicators in the Near-infrared (SVADhIN) program, which is aimed at obtaining homogeneous (and improved) distance measurements to a number of nearby ($D<150$~kpc) classical dSph galaxies, via time-series near-infrared (NIR) observations of their known RR Lyrae and other pulsating stars. This is because RR Lyrae follow tight Period-Luminosity Relations (PLRs) at NIR wavelengths \citep{marconi2015,bhardwaj2023}, allowing the possibility to derive $\sim1\%$ accurate and precise RR Lyrae based distances to nearby galaxies. In particular, \citet{bhardwaj2023} provided the most precise empirical calibration of Period-Luminosity-Metallicity Relations (PLZRs) at NIR wavelengths using more than 1300 variables in 11 globular clusters and Milky Way field. The NIR photometry has several advantages over optical data for RR Lyrae distance measurements, in particular, tighter PLRs, smaller extinction, smaller amplitude and temperature variations \citep[for examples, see][]{longmore1986,bono2001,catelan2004,sollima2006,braga2015,muraveva2015,beaton2018,bhardwaj2020, bhardwaj2022}. 

The first paper in the series presented results on Draco dSph in \citet[][hereafter Paper I]{bhardwaj2024}, reaching a precision of $\sim 1.5\%$ on its distance based on the time-series NIR observations using the WIRCam \citep{puget2004} mounted on the Canada France Hawaii Telescope (CFHT). The CFHT/WIRCam observations of a few additional nearby classical dSph are still on-going. As CFHT can only observe classical dSph in the northern sky, we also targeted a few southern dSph, for example, the Carina dSph using the Magellan Telescope in this work. For a detailed summary of studies on Carina dSph, the readers are referred to \citet{deboer2014}, \citet{santana2016}, and \citet{norris2017}. 

In this manuscript, Section \ref{sec_obs} describes Magellan observations and the associated photometric data in Section \ref{sec_phot}. We present pulsating stars detected in our dataset including RR Lyrae, anomalous Cepheids (ACep) and dwarf Cepheids (DCep) in Section \ref{sec_var}. Our results on distance moduli to Carina dSph using the empirical NIR PLRs and/or PLZRs are presented in Section \ref{sec_mu0}, followed by discussion and conclusions in Section \ref{sec_end}.

\section{Observations} \label{sec_obs}

\begin{figure}
  \epsscale{1.1}
  \plotone{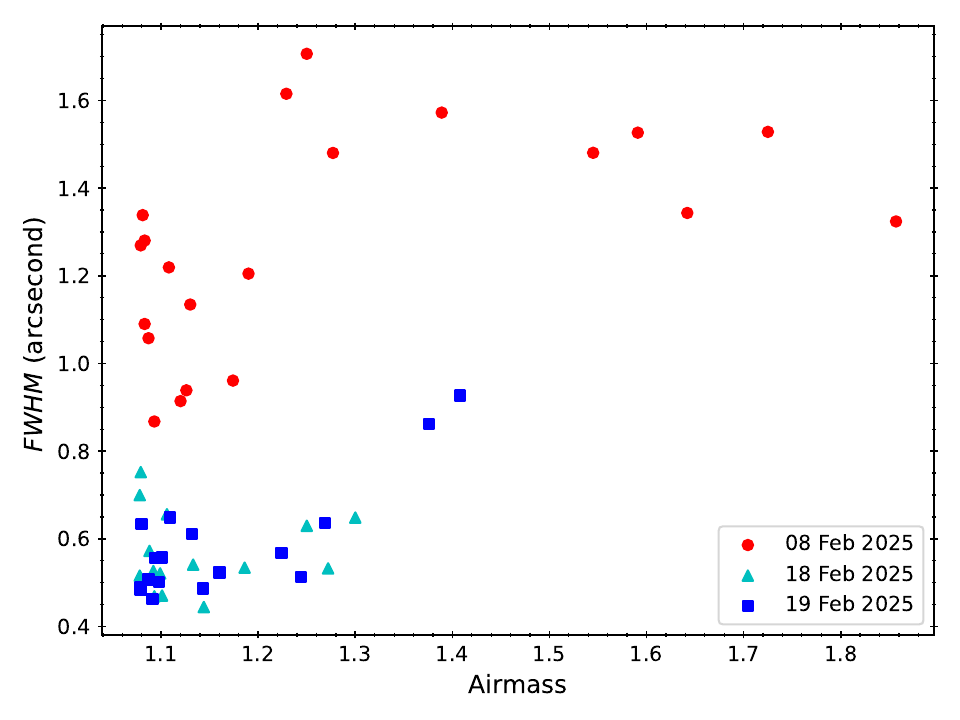}
  \caption{Seeing as a function of airmass for the Magellan/FourStar observations.} 
  \label{fig_seeing}
\end{figure}

Time-series NIR observations of Carina dSph were carried out using the FourStar infrared camera \citep{persson2013} mounted on one of the 6.5-meter Magellan (Baade) Telescope. The FourStar camera consists of an array of HAWAII-2RG detectors in $2\times 2$ layout, each of them have $2048 \times 2048$ pixels, providing a pixel scale of $0.159\arcsec$ per pixel and a total field-of-view (FOV) of $\sim 10.8\arcmin \times 10.8\arcmin$. The remote time-series observations were done on the full night of 08 February 2025, and two first-half nights on 18-19 February 2025. As can be seen from Figure \ref{fig_seeing}, the weather on the first night was sub-optimal with a median seeing of $\sim 1.3\arcsec$. In contrast, observing conditions in other two half nights were excellent, achieving a median seeing of $\sim 0.5\arcsec$ on both nights. 

\begin{figure}
  \epsscale{1.2}
  \plotone{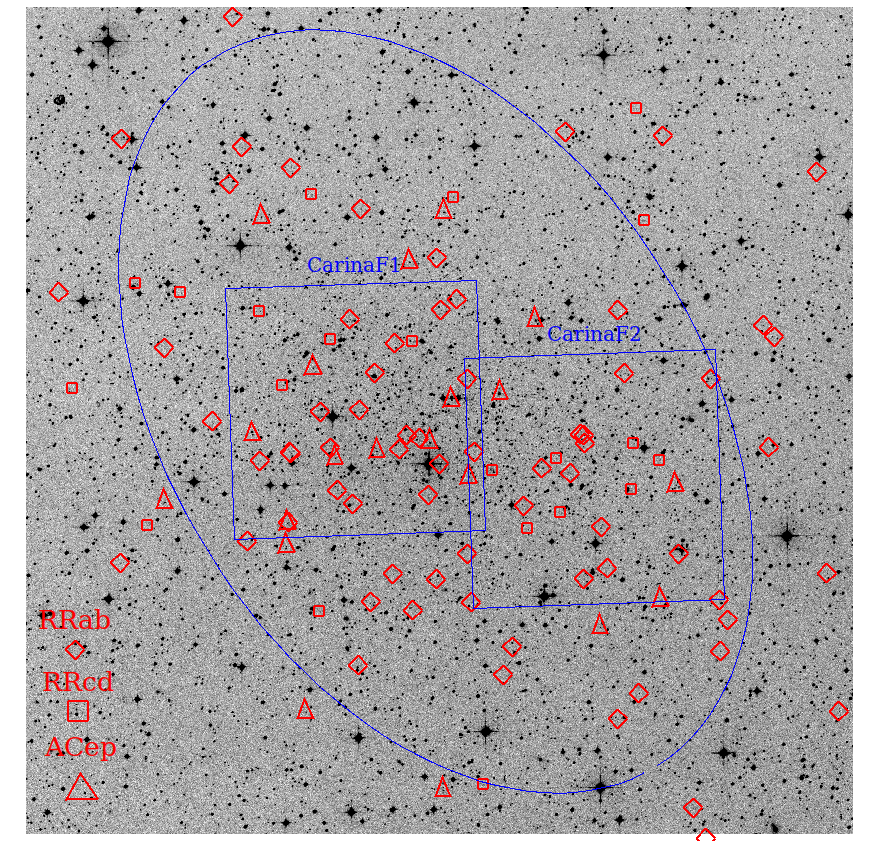}
  \caption{Spatial distribution of two fields of Carina dSph, observed by Magellan/FourStar, overlaid on the Digitized Sky Survey (DSS) red plates image (with a size of $40\arcmin \times 40\arcmin$). The blue ellipse represents the contour with semimajor axis twice of the half-light radius ($r_h$) of Carina dSph, using the following parameters adopted from \citet{munoz2018}: $r_h = 10.1\arcmin$, ellipticity of 0.36, and position angle of $60^\circ$. The diamonds, squares and triangles are the locations of fundamental mode RR Lyrae (RRab), first-overtone and double mode RR Lyrae (RRcd), and ACep, respectively. For clarity, locations of DCep were omitted.}
  \label{fig_cov}
\end{figure}

Since Carina dSph is quite extended, with a half-light radius of $10.1\arcmin$ \citep{munoz2018}, our FourStar observations were centered on two fields: CarinaF1 at $\alpha_{J2000}=$06:42:02.1, $\delta_{J2000}=-$50:57:20.4, and CarinaF2 at $\alpha_{J2000}=$06:40:51.4, $\delta_{J2000}=-$51:01:12.5 (see Figure \ref{fig_cov}). Each field was observed in the $JHK_S$ sequence, alternating multiple times between both fields in a given night. As the goal of these observations was to obtain time-series data of a subset of RR Lyrae located in the Carina dSph, we did not intend to increase the number of FourStar fields to cover the larger fraction of the galaxy within the allocated nights.

For a single epoch $J$- and $H$-band observations, we executed a 5-point dithering sequence to properly subtract the rapidly varying NIR sky background and to cover the $\sim 19\arcsec$ gaps between the detectors. This dithering sequence was repeated 10 times to achieve the desired signal-to-noise ratio (SNR). In the case of $K_s$-band, we carried out a 11-point dithering pattern and repeated it 20 times for deep observations. Each frame in the dithering pattern had identical exposure time of $5.822$~second, resulting a total observing time of $\sim 7.8$~minutes in the $JH$-band and $\sim33.2$~minutes in the $K_s$-band for a single epoch observation.\footnote{The actual total exposure time is 4.85~min and 21.34~min in the $JH$- and $K_s$-band, respectively.} According to the FourStar exposure time calculator, this observing strategy allowed us to reach a SNR$>15$ for the typical brightness of RR Lyrae in Carina dSph (about $20$~mag in $K_s$-band). Therefore, including the overheads, a sequence of $JHK_S$ observation in a given Carina field took about an hour. In total, we collected 9 epochs of $JHK_s$ data on the two Carina fields, with an exception of the $K_s$-band data for CarinaF2 field where only 8 epochs were observed. All of the imaging data were reduced using the dedicated FSRED FourStar reduction pipeline,\footnote{\url{https://instrumentation.obs.carnegiescience.edu/FourStar/SOFTWARE/reduction.html}} which also included a set of dark and dome-flat images taken on the same nights. 

\section{Photometry and the Color-Magnitude Diagram} \label{sec_phot}

\begin{figure}
  \epsscale{1.2}
  \plotone{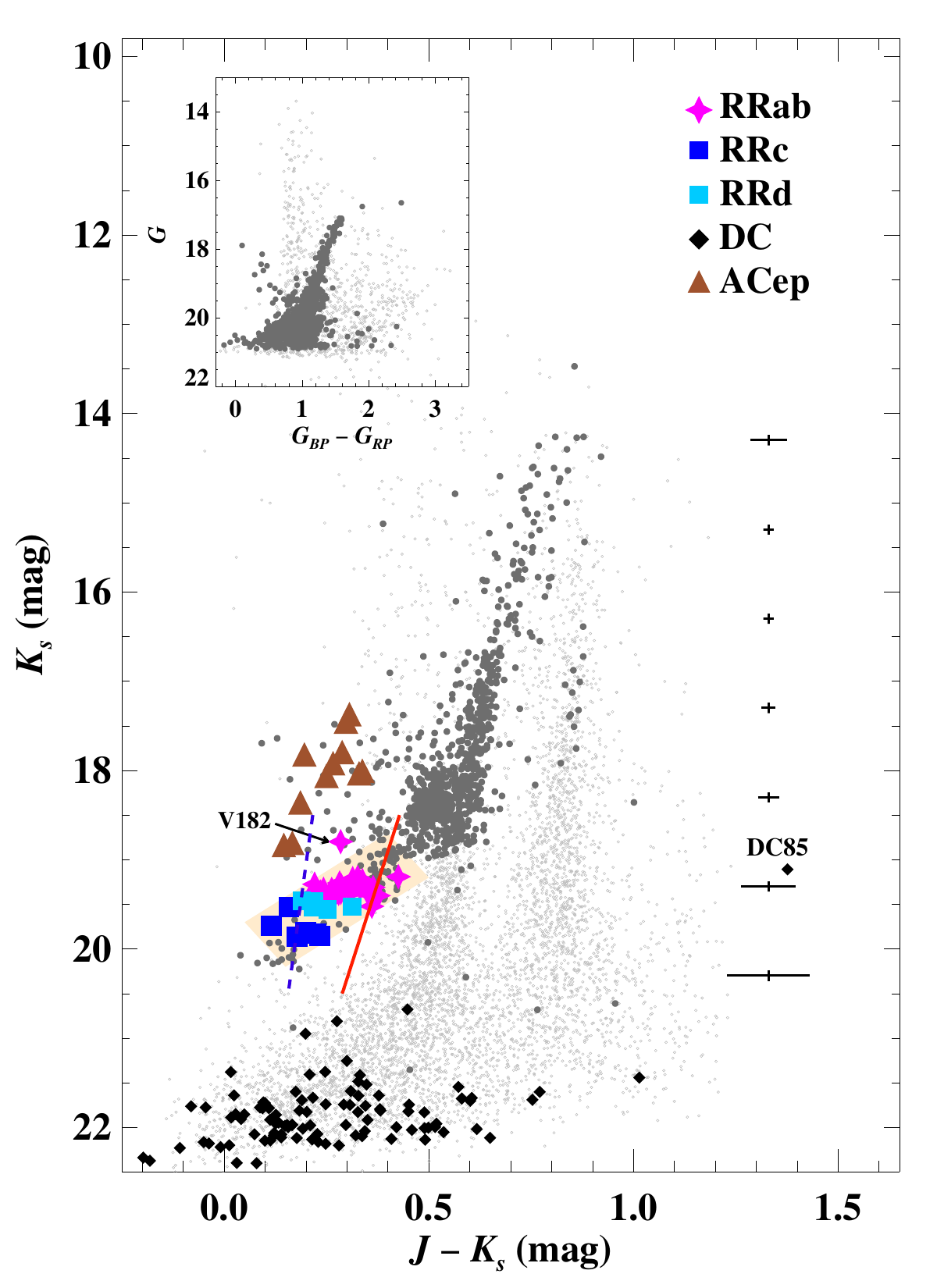}
  \caption{Extinction corrected CMD in the NIR for the combined sources (light grey dots) in the two Carina fields. The dark grey circles represent the sources with membership probability greater than 0.5, as determined by \citet{battaglia2022}. Foreground contamination can be clearly seen in the CMD as a vertical stripe of sources around $(J-K_S)\sim 0.8$ that are mostly the late-type dwarf stars in the Milky Way. Variable stars studied in this work are overplotted with their mean magnitudes and colors; clear outliers (V182 and DC85) are explained in the text. The theoretically predicted instability strip boundaries - fundamental red edge (solid red line) and first-overtone blue edge (dashed blue line) are taken from \citet{marconi2015}, shifted using a distance modulus of $\sim 20.07$~mag \citep[an averaged value based on Table 1 of][]{karczmarek2015}. The inset figure represents the CMD in the Gaia filters for sources cross-matched to the \citet{battaglia2022} catalog.
  On the right the representative $\pm3\sigma$ error bars are shown as a function of magnitude and color.}
  \label{fig_cmd}
\end{figure}

\begin{figure*}
  \epsscale{1.1}
  \plottwo{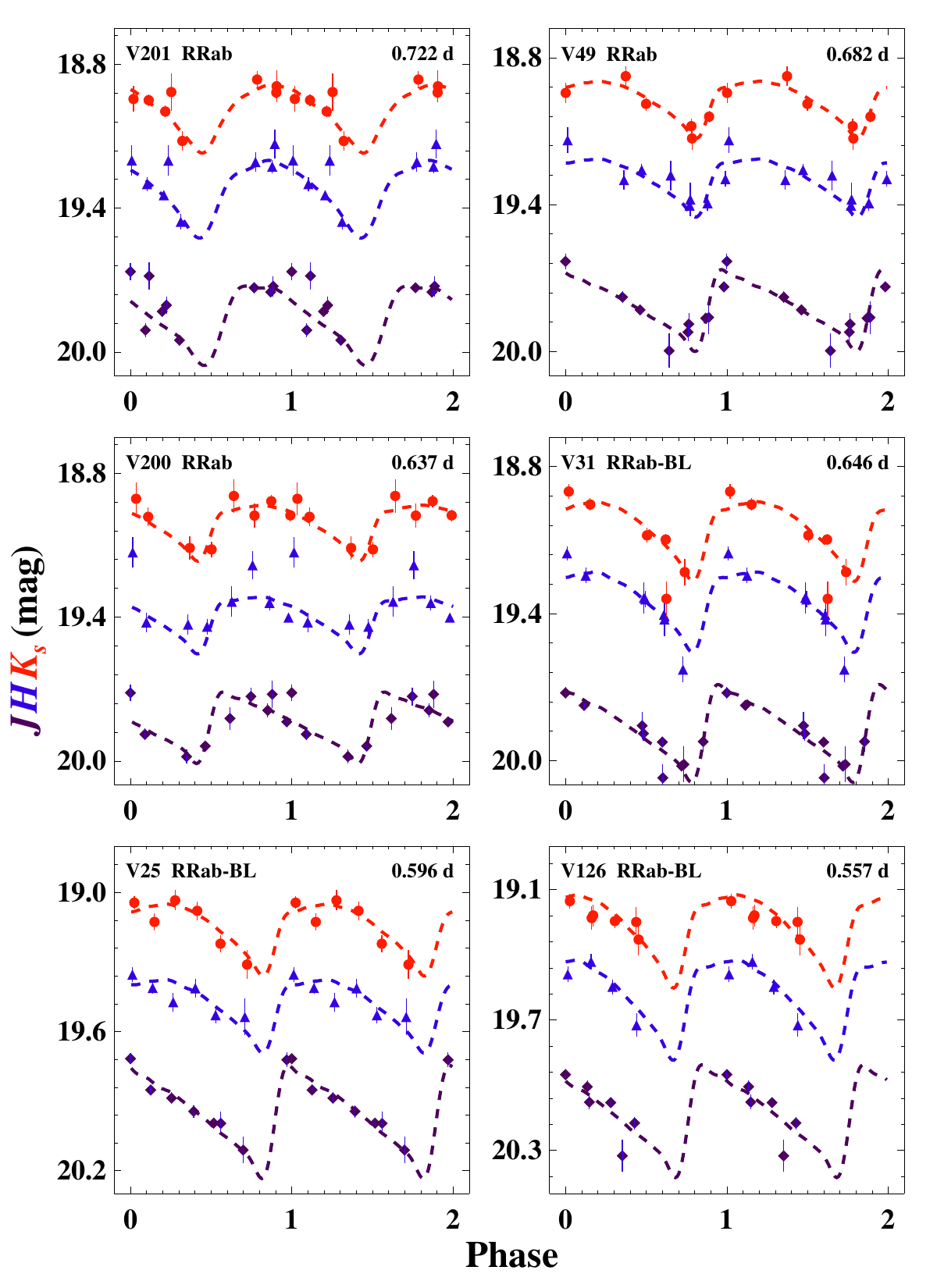}{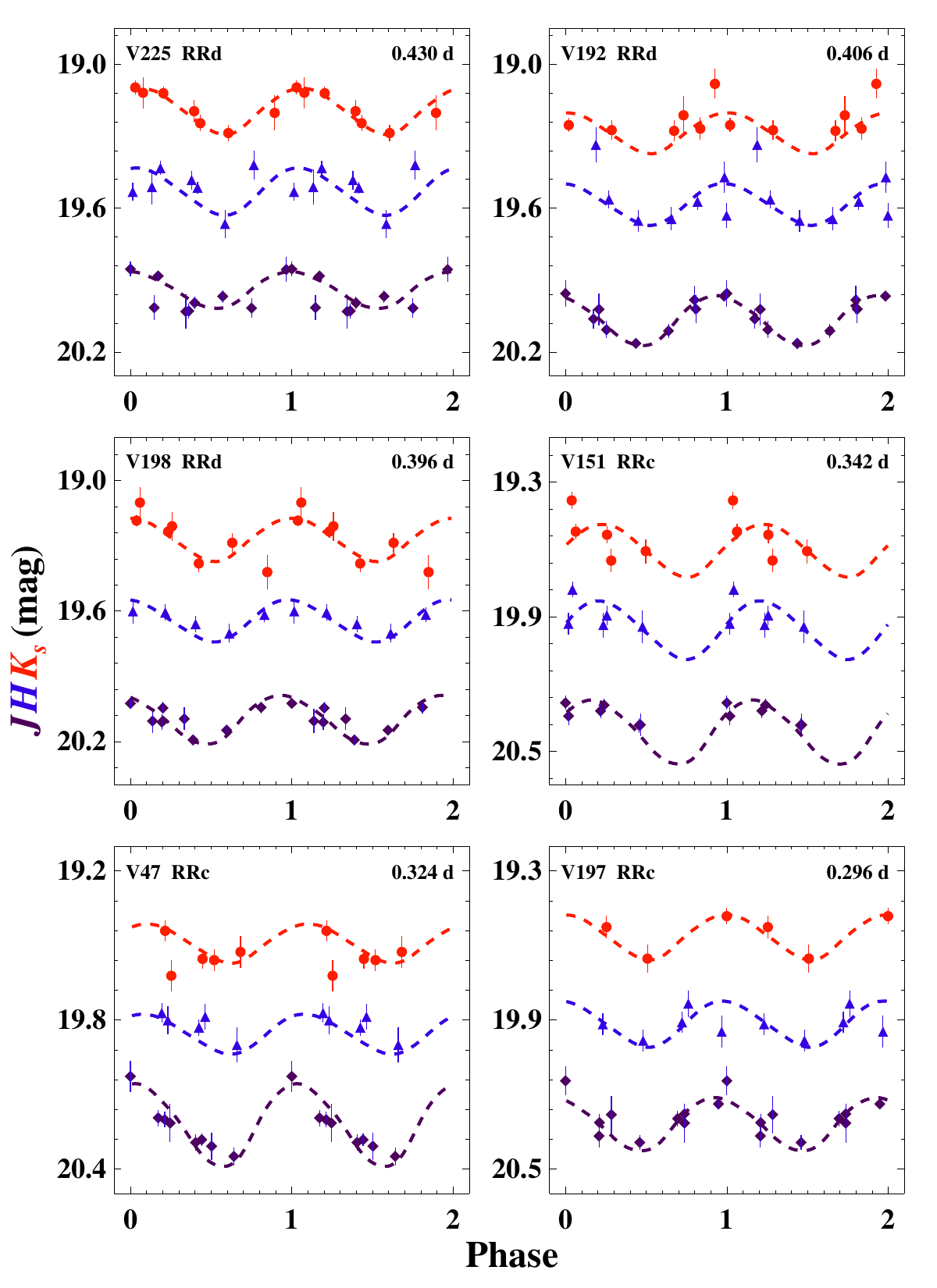}
  \caption{Examples of the phased light curves in the $JHK_S$ bands for RR Lyrae. {\it Left:} RRab stars. {\it Right:} RRc/RRd stars. For a better visualisation, the $J$- and $K_S$-band light curves have been shifted by $+0.2$~mag and $-0.3$~mag, respectively. The dashed lines represent the best-fit templates. The variable star ID, subtype, and the pulsation period are included at the top of each panel. The first-overtone mode periods were used to phase light curves of RRd stars.}
  \label{fig_rr}
\end{figure*}

The point-spread-function (PSF) photometry was performed on all of the processed epoch images as well as on the deep stacked images. The photometry was done using the standard {\tt DAOPHOT/ALLSTAR} \citep{1987PASP...99..191S} and {\tt ALLFRAME} \citep{1994PASP..106..250S} suite of softwares as described in detail in Paper I (and reference therein). Finally, the PSF photometry was calibrated to the Two Micron All Sky Survey \citep[2MASS,][]{2006AJ....131.1163S} using the 2MASS point-source catalog \citep{2003yCat.2246....0C}. Along with the calibrated $JHK_s$ photometry, the light curves for each candidate variable star were also generated by correcting for the epoch-dependent frame-to-frame zero-point variations, as described in Paper I.

In total, about $10^4$ sources were detected from the deep stacked images in both fields. Figure \ref{fig_cmd} presents the NIR color-magnitude diagram (CMD) by merging the sources in both Carina fields. Individual sources in this CMD have been corrected for the foreground extinction using the \citet[][via the {\tt dustmaps} code]{schlegel1998} dust maps and using $A_{J,H,K_S} = \{0.94,\ 0.58,\ 0.39\}\times E(B-V)$ (for details, see Paper I). Small variations in the adopted extinction law and color-excess values do not add any significant uncertainty in the near-infrared magnitudes.

Based on the Gaia early Data Release 3 (EDR3) data, \citet{battaglia2022} determined the membership probabilities for 74 dwarf galaxies including Carina dSph. Therefore, we cross-matched our catalog to their list of membership. Figure \ref{fig_cmd} highlights the sources with membership probability greater than 0.5, which clearly trace the red-giant branch and red-clump stars in Carina dSph. Since Carina dSph is also located at a low Galactic latitude ($b=-22.2^\circ$) and is quite extended, contamination from foreground dwarf stars is unavoidable \citep[for examples, see][]{bono2010,santana2016}. The variable horizontal branch is well populated with most stars occupying the location within the predicted instability strip boundaries in \citet{marconi2015}.

\section{Pulsating Stars} \label{sec_var}

Dwarf spheroidal galaxies host a number of variable stars including pulsating stars such as RR Lyrae, Anomalous Cepheids (ACep), and Dwarf Cepheids (DCep) \citep[see review by][]{monelli2022}. RR Lyrae are old-population pulsators and ACep are generally associated with intermediate-age populations. DCep could refer to the population-II SX Phoenicis (SX Phe) stars, however sometimes they also include the population I $\delta$-Scuti (DSCT) stars. These  stars primarily exhibit radial mode pulsations, except in DSCT stars which show both radial and non-radial modes. These pulsating stars are located within the instability strip on the CMD, where ACep are the brightest pulsators followed by RR Lyrae and DCep. Similarly, DCep and ACep have the shortest and longest pulsation periods ($\sim 1-2$~hours, and $\sim 0.5-2.5$~days, respectively), and RR Lyrae cover a period range from $\sim 0.3$ to $\sim 1$~days.

Based on photographic plate observations, \citet{saha1986} discovered 172 potential variable stars\footnote{Some of them turned out to be a non-variable star \citep{mateo1998,dallora2003,coppola2015}.} toward Carina dSph, with 58 of them identified as RR Lyrae belonging to Carina. A few other variable stars were later identified as ACep \citep{dacosta1988,nemec1988}. Using CCD observations, \citet{dallora2003} recovered most of the RR Lyrae and ACep, together with 26 and 7 new discoveries, respectively. Hence, the number of these two types of pulsating stars was increased to 90 in Carina dSph. These early observations were restricted to $BV$ bands and only observed in three to four consecutive nights. On the other hand, multi-site observations done in \citet{coppola2013} and \citet{coppola2015} extended the filters coverage to $UBV(R)I$ and spanned more than 20 years, resulting in a comprehensive catalog consisting of 92 RR Lyrae and 20 ACep in Carina dSph.

\begin{deluxetable*}{lccccccccc}
  \tabletypesize{\scriptsize}
  \tablecaption{Mean $JHK_S$-band photometry for Anomalous Cepheids and RR Lyrae in Carina dSph.}
  \label{tab_meanmag}
  \tablewidth{0pt}
  \tablehead{
    \colhead{ID} &
    \colhead{Type} &
    \colhead{$P$ [days]} &
    \colhead{$J$ [mag]} &
    \colhead{$H$ [mag]} &
    \colhead{$K_S$ [mag]} &
    \colhead{$N_J$\tablenotemark{a}} &
    \colhead{$N_H$\tablenotemark{a}} &
    \colhead{$N_{K_S}$\tablenotemark{a}} &
    \colhead{$E(B-V)$ [mag]} 
  }
  \startdata
  \multicolumn{10}{c}{Anomalous Cepheids} \\
V41   & F	& 1.038153 & $18.193\pm0.078$ & $17.946\pm0.056$ & $17.920\pm0.060$ & 9/9 & 9/8 & 9/9 & 0.065 \\
V87   & FO	& 0.855613 & $18.030\pm0.030$ & $17.828\pm0.032$ & $17.820\pm0.019$ & 9/9 & 9/8 & 9/9 & 0.060 \\
V149  & F	& 0.917707 & $19.030\pm0.107$ & $18.802\pm0.084$ & $18.848\pm0.083$ & 8/8 & 8/8 & 7/7 & 0.051 \\
V187  & F	& 0.950292 & $18.286\pm0.046$ & $18.085\pm0.029$ & $18.045\pm0.049$ & 9/9 & 9/9 & 8/7 & 0.051 \\
V190  & F	& 1.164710 & $18.112\pm0.058$ & $17.861\pm0.027$ & $17.843\pm0.021$ & 9/9 & 9/9 & 9/9 & 0.063 \\
V193  & FO	& 0.426358 & $18.592\pm0.021$ & $18.407\pm0.017$ & $18.402\pm0.025$ & 9/9 & 9/8 & 9/9 & 0.061 \\
V203  & F	& 0.939887 & $18.402\pm0.038$ & $18.109\pm0.030$ & $18.068\pm0.038$ & 9/9 & 9/8 & 9/8 & 0.066 \\
V205  & FO	& 0.383368 & $19.051\pm0.024$ & $18.877\pm0.044$ & $18.850\pm0.023$ & 8/8 & 8/7 & 8/8 & 0.070 \\
V216  & FO	& 1.079230 & $17.787\pm0.024$ & $17.413\pm0.022$ & $17.442\pm0.025$ & 9/8 & 9/9 & 8/8 & 0.057 \\
V219  & FO	& 1.365517 & $17.765\pm0.017$ & $17.477\pm0.012$ & $17.437\pm0.013$ & 9/9 & 9/8 & 9/9 & 0.067 \\
V230  & F	& 1.002573 & $18.188\pm0.030$ & $17.970\pm0.020$ & $17.955\pm0.027$ & 9/9 & 9/8 & 9/9 & 0.066 \\
\multicolumn{10}{c}{RR Lyrae} \\
V25  & RRab-BL	& 0.595651 & $19.751\pm0.057$ & $19.478\pm0.028$ & $19.447\pm0.047$ & 9/9 & 8/8 & 9/9 & 0.067 \\
V30  & RRab	& 0.618811 & $19.742\pm0.060$ & $19.456\pm0.039$ & $19.427\pm0.035$ & 8/8 & 8/8 & 8/7 & 0.070 \\
V31  & RRab-BL	& 0.645795 & $19.659\pm0.039$ & $19.303\pm0.052$ & $19.339\pm0.069$ & 9/8 & 9/9 & 9/8 & 0.068 \\
V32  & RRab	& 0.615052 & $19.679\pm0.034$ & $19.437\pm0.045$ & $19.387\pm0.022$ & 9/9 & 8/8 & 9/9 & 0.068 \\
V44  & RRab	& 0.626428 & $19.692\pm0.041$ & $19.443\pm0.066$ & $19.361\pm0.046$ & 9/9 & 8/8 & 9/9 & 0.066 \\
V49  & RRab	& 0.681507 & $19.608\pm0.035$ & $19.291\pm0.030$ & $19.270\pm0.034$ & 9/9 & 9/9 & 9/8 & 0.068 \\
V57  & RRab	& 0.612902 & $19.757\pm0.047$ & $19.471\pm0.040$ & $19.431\pm0.050$ & 9/9 & 9/9 & 8/8 & 0.064 \\
V58  & RRab	& 0.619460 & $19.637\pm0.062$ & $19.255\pm0.070$ & $19.320\pm0.062$ & 9/9 & 8/8 & 8/7 & 0.065 \\
V65  & RRab	& 0.651711 & $19.684\pm0.046$ & $19.385\pm0.067$ & $19.368\pm0.036$ & 9/8 & 8/7 & 9/9 & 0.064 \\
V68  & RRab	& 0.678736 & $19.585\pm0.023$ & $19.317\pm0.030$ & $19.233\pm0.033$ & 9/9 & 9/7 & 9/9 & 0.065 \\
V75  & RRab	& 0.591216 & $19.763\pm0.041$ & $19.464\pm0.044$ & $19.426\pm0.060$ & 9/9 & 9/9 & 9/8 & 0.063 \\
V84  & RRab	& 0.616681 & $19.681\pm0.037$ & $19.379\pm0.025$ & $19.401\pm0.032$ & 9/7 & 9/9 & 9/9 & 0.063 \\
V90  & RRab	& 0.631363 & $19.663\pm0.056$ & $19.409\pm0.050$ & $19.314\pm0.062$ & 9/8 & 8/7 & 9/8 & 0.058 \\
V116  & RRab	& 0.683313 & $19.726\pm0.033$ & $19.462\pm0.038$ & $19.404\pm0.025$ & 9/9 & 9/9 & 8/8 & 0.056 \\
V122  & RRab	& 0.631469 & $19.637\pm0.024$ & $19.366\pm0.041$ & $19.353\pm0.040$ & 9/9 & 9/8 & 7/7 & 0.055 \\
V124  & RRab	& 0.591721 & $19.709\pm0.029$ & $19.407\pm0.023$ & $19.261\pm0.022$ & 9/9 & 9/9 & 8/8 & 0.055 \\
V125  & RRab	& 0.594096 & $19.706\pm0.056$ & $19.406\pm0.054$ & $19.403\pm0.027$ & 9/8 & 9/9 & 8/8 & 0.053 \\
V126  & RRab-BL	& 0.557097 & $19.923\pm0.062$ & $19.569\pm0.051$ & $19.557\pm0.028$ & 9/9 & 9/9 & 7/7 & 0.053 \\
V127  & RRab-BL	& 0.626295 & $19.707\pm0.035$ & $19.439\pm0.033$ & $19.402\pm0.044$ & 9/8 & 9/9 & 8/8 & 0.053 \\
V133  & RRab	& 0.612313 & $19.795\pm0.031$ & $19.538\pm0.036$ & $19.520\pm0.034$ & 9/9 & 9/9 & 8/7 & 0.053 \\
V135  & RRab	& 0.590919 & $19.636\pm0.040$ & $19.421\pm0.030$ & $19.428\pm0.031$ & 9/7 & 9/9 & 8/8 & 0.053 \\
V141  & RRab	& 0.635334 & $19.662\pm0.038$ & $19.377\pm0.039$ & $19.368\pm0.031$ & 9/9 & 9/7 & 8/8 & 0.053 \\
V153  & RRab	& 0.660369 & $19.637\pm0.022$ & $19.356\pm0.035$ & $19.375\pm0.037$ & 9/9 & 9/8 & 8/8 & 0.050 \\
V179  & RRab	& 0.663794 & $19.615\pm0.016$ & $19.320\pm0.018$ & $19.280\pm0.033$ & 9/9 & 9/9 & 9/9 & 0.062 \\
V182  & RRab	& 0.788972 & $19.108\pm0.026$ & $18.800\pm0.016$ & $18.891\pm0.022$ & 9/9 & 9/9 & 9/9 & 0.059 \\
V183  & RRab	& 0.612302 & $19.586\pm0.019$ & $19.338\pm0.024$ & $19.295\pm0.032$ &16/16&17/15&14/13& 0.058 \\
V191  & RRab	& 0.650289 & $19.598\pm0.034$ & $19.343\pm0.024$ & $19.255\pm0.040$ &17/17&15/13&14/13& 0.060 \\
V195  & RRab	& 0.618968 & $19.764\pm0.038$ & $19.466\pm0.030$ & $19.417\pm0.034$ & 9/8 & 9/7 & 9/8 & 0.065 \\
V196  & RRab	& 0.665989 & $19.599\pm0.030$ & $19.320\pm0.031$ & $19.303\pm0.036$ & 9/8 & 9/8 & 8/8 & 0.058 \\
V200  & RRab	& 0.637347 & $19.640\pm0.035$ & $19.393\pm0.041$ & $19.307\pm0.028$ & 9/9 & 8/6 & 9/9 & 0.064 \\
V201  & RRab	& 0.722244 & $19.649\pm0.037$ & $19.320\pm0.046$ & $19.290\pm0.023$ & 9/9 & 8/7 & 9/8 & 0.067 \\
V204  & RRab	& 0.633269 & $19.646\pm0.062$ & $19.395\pm0.034$ & $19.321\pm0.036$ & 8/8 & 8/8 & 9/9 & 0.069 \\
V47  & RRc	& 0.323767 & $20.011\pm0.036$ & $19.854\pm0.023$ & $19.787\pm0.023$ & 9/9 & 9/9 & 9/9 & 0.065 \\
V142  & RRc	& 0.363543 & $19.953\pm0.023$ & $19.702\pm0.030$ & $19.766\pm0.051$ & 9/9 & 8/8 & 8/8 & 0.052 \\
V144  & RRc	& 0.393357 & $19.753\pm0.018$ & $19.523\pm0.036$ & $19.535\pm0.021$ & 9/9 & 9/9 & 8/7 & 0.052 \\
V151  & RRc	& 0.341801 & $20.205\pm0.023$ & $19.954\pm0.027$ & $19.898\pm0.041$ & 9/8 & 9/7 & 8/8 & 0.052 \\
V181  & RRc	& 0.279491 & $20.123\pm0.023$ & $19.977\pm0.046$ & $19.931\pm0.052$ & 9/9 & 8/8 & 9/9 & 0.066 \\
V197  & RRc	& 0.296172 & $20.113\pm0.021$ & $19.913\pm0.025$ & $19.862\pm0.043$ & 9/9 & 9/8 & 8/8 & 0.056 \\
V26  & RRd	& 0.562177\tablenotemark{b} & $19.734\pm0.032$ & $19.459\pm0.027$ & $19.428\pm0.024$ & 9/9 & 8/8 & 9/9 & 0.063 \\
V74  & RRd	& 0.533702\tablenotemark{b} & $19.792\pm0.043$ & $19.616\pm0.043$ & $19.562\pm0.061$ & 9/9 & 9/9 & 9/8 & 0.061 \\
V192  & RRd	& 0.541694\tablenotemark{b} & $19.844\pm0.041$ & $19.550\pm0.059$ & $19.571\pm0.018$ & 9/9 & 9/9 & 8/7 & 0.059 \\
V198  & RRd	& 0.530551\tablenotemark{b} & $19.896\pm0.034$ & $19.660\pm0.055$ & $19.521\pm0.076$ & 9/8 & 9/9 & 8/8 & 0.059 \\
V225  & RRd	& 0.576880\tablenotemark{b} & $19.809\pm0.031$ & $19.521\pm0.051$ & $19.463\pm0.048$ & 9/9 & 9/9 & 8/8 & 0.055 \\
\enddata
\tablenotetext{a}{The two numbers respectively represent the number of detections from the single-epoch images and the final number of data-points used to determine the mean magnitudes (see text for more details).}
\tablenotetext{b}{The fundamental-mode periods for RRd were converted to the dominant first-overtone periods using the period ratios given in \citet{coppola2015} for light curve phasing.}
\end{deluxetable*}

\begin{figure*}
  \epsscale{1.0}
  \plotone{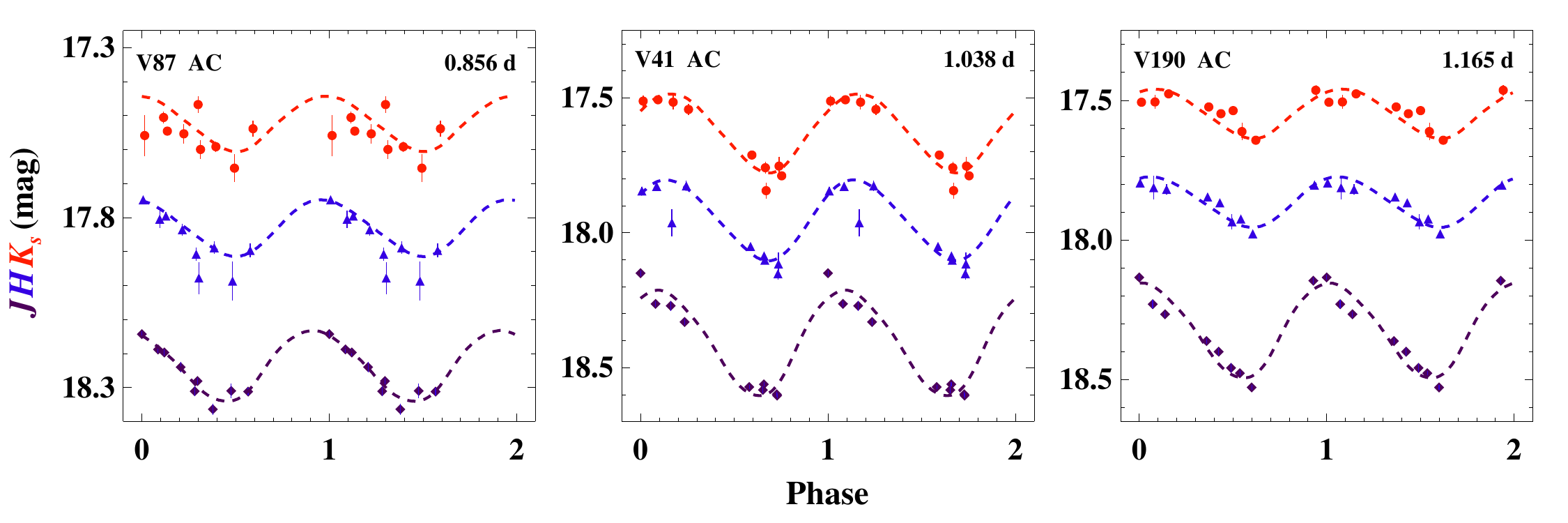}
  \caption{Same as Figure \ref{fig_rr}, but for the three representative ACep pulsating stars.}
  \label{fig_ac}
\end{figure*}

\begin{figure*}
  \epsscale{1.1}
  \plotone{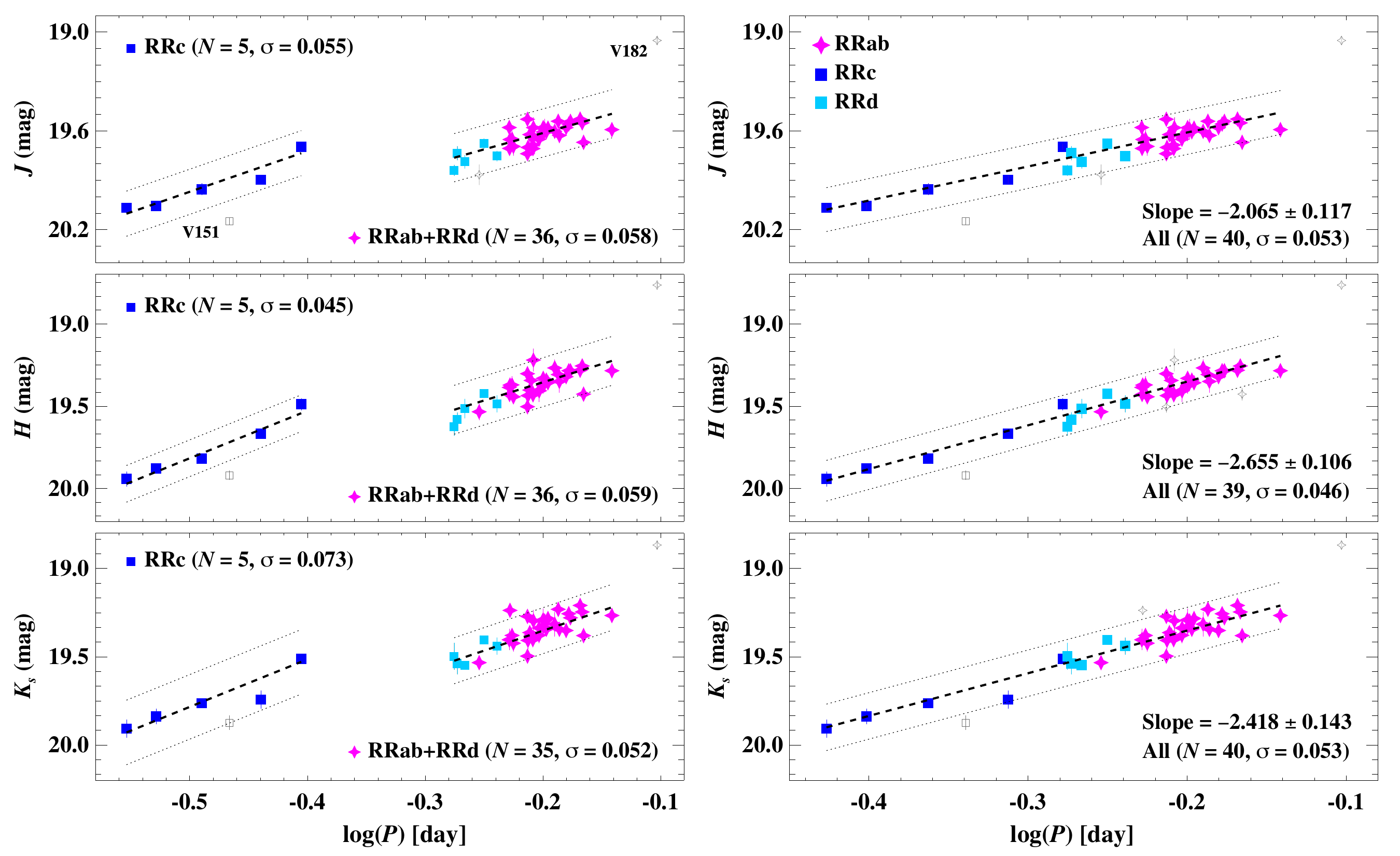}
  \caption{The extinction-corrected NIR period-luminosity relations for RRab+d and RRc stars (left) and all RRL stars (right) in $J$ (top), $H$ (middle), and $K_s$ (bottom). In the left panels, RRd stars were used with their fundamental-mode periods from \citet{coppola2015}. In the right panels, the periods of both the RRc and RRd stars have been shifted to their corresponding fundamental-mode periods, as explained in the text. The grey symbols represent outliers removed iteratively during the fitting procedure. The dashed lines represent best-fitting linear regressions over the period range under consideration, while the dotted lines display $\pm 2.5\sigma$ offsets from the best-fitting PLRs.} 
  \label{fig_rrpl}
\end{figure*}

At the same time, \citet{vivas2013} surveyed a larger area, out to $\sim3\times$ of the tidal radius along the semimajor axis of Carina dSph. Since their observing strategy was designed to search for DCep, they only recovered 30 RR Lyrae reported in \citet{dallora2003}, but with 8 new discoveries. Similarly, they recovered 6 ACep from \citet{dallora2003} and discovered 4 new ones. Most of the newly discovered RR Lyrae and ACep were located far from the center of Carina dSph. \citet{vivas2013} discovered a large number of 340 DCep in Carina dSph. They also confirmed 12 DCep candidates listed in \citet{mateo1998}, who discovered the first 20 DCep in Carina dSph using modern CCD observations. This large sample of DCep was further augmented by 101 new discoveries presented in \citet{coppola2015}, bringing the total number of DCep in Carina dSph close to $\sim 440$. Combining the above catalogs, there are 43 RR Lyrae, 11 ACep, and 102 DCep that fall within the two Carina fields in our observations. All of them are included in the \citet{coppola2015} catalog, and we adopted their ID, periods, and variable stars classification for our analysis of pulsating stars.

\subsection{Mean Magnitudes and RR Lyrae PL relations} \label{sec_meanmag}

The calibrated $JHK_S$-band light curves for the RR Lyrae and ACep identified in our FourStar fields were fitted using the template light curves presented in \citet{braga2019} and sinusoidal templates, respectively, as done in Paper I, to derive their mean $JHK_s$-band magnitudes. Since some of the outliers can be seen on the light curves (e.g. V200), mainly from the first night of observations with the worst seeing (see Figure \ref{fig_seeing}), we removed the outliers using a recursive $2.5\sigma$-rejection algorithm while fitting the template light curves. Examples of light curves, together with the best-fit template light curves, are shown in Figures \ref{fig_rr} and \ref{fig_ac} for RR Lyrae and ACep, respectively. There are two RR Lyrae, V183 and V191, located near the edges of both fields. These were observed twice compared to other RR Lyrae. For double-mode RRd stars, we converted their fundamental-mode (F) periods to the first-overtone (FO) periods using the period ratios listed in \citet{coppola2015}, as the latter resulted in more sinusoidal light curve which in turn allowed the mean magnitudes to be determined using templates. The locations of these pulsating stars in the CMD are shown in Figure \ref{fig_cmd} based on the mean magnitudes listed in Table \ref{tab_meanmag}.    

The fainter DCep, on the other hand, required a different approach for the mean-magnitude determination. This is because the pulsation period for DCep is around $\sim 1$ to $\sim 2$~hours, which is comparable to the time taken for a sequence of $JHK_S$ observation (about an hour, see Section \ref{sec_obs} for details). Furthermore, DCep are intrinsically fainter than RR Lyrae by $\sim 1.5$ to $\sim 2$~mag, and thus their observed brightness with $K_s\sim 22$~mag is close to the detection limit of our observation (see Figure \ref{fig_cmd}). Thus, most of those are either undetected (especially from the first night) or carried a large photometric error in single-epoch images. Therefore, DCep photometry was obtained from the deep stacked images (the same images used to create the CMD shown in Figure \ref{fig_cmd}), and we adopted these photometric measurements as mean DCep magnitudes.  

We derived PLRs for different subsets of RR Lyrae stars i.e. RRab, RRc, and the combined sample of all RR Lyrae stars. The combined sample (ab+cd) provides an extended period-range and increased statistics to better constrain the slope and intercept of the PLRs. All of the mean magnitudes have been corrected for extinction prior to the fitting of PLRs, with $E(B-V)$ values determined from the \citet{schlegel1998} dust maps. When combining the different samples of RR Lyrae, we fundamentalized the RRc periods using $\log P_F = \log P_O + 0.127$ \citep[see Paper I and][]{coppola2015}, where $P_F$ and $P_O$ represent the F and FO periods, respectively. With this fundamentalization equation, no obvious offsets were found in the residuals of combined PLRs in our previous works \citep[][and Paper I]{bhardwaj2020a}. Considering a narrow range in period distribution of RRab stars, we used F periods of RRd stars from \citet{coppola2015} to fit the linear regression to the combined RRab and RRd sample.  Figure~\ref{fig_rrpl} shows NIR PLRs for RR Lyrae stars in Carina dSph and results of the linear regression are listed in Table~\ref{tbl:plr_rr}.  Among the sample of 43 RR Lyrae, V182 is a clear outlier as can be seen in the CMD. An RRc variable, V151 is systematically fainter than the PLR plane. The slopes of the combined sample of RR Lyrae stars are in good agreement with those for Draco RR Lyrae in Paper I. 

While fitting the PLRs, we adopted a $2.5\sigma$-clipping algorithm to recursively remove outliers. Small statistics (specially for RRc) and a narrow period-range of RR Lyrae means more sensitivity to potential outlier(s) or even slightly uncertain magnitudes. This stricter threshold provides robust measurements of slopes and intercepts and enables a relative comparison with results of Paper I. No statistically significant changes are seen in the results with the the variation in adopted $\sigma$ clipping for outlier removal.

\begin{deluxetable}{cccccc}
\tablecaption{Near-infrared PL relations of RRL stars in Carina dSph. \label{tbl:plr_rr}}
\tablewidth{0pt}
\tablehead{
{Band} & {Type} & {$a_\lambda$} & {$b_\lambda$} & {$\sigma$}& {$N$}}
\startdata
     $J$ &  RRab &    19.215$\pm$0.065      &$     -1.989\pm0.319      $&      0.058 &   36\\
     &   RRc &    18.734$\pm$0.155      &$     -2.473\pm0.325      $&      0.055 &    5\\
     &   All &    19.197$\pm$0.029      & $    -2.065\pm0.117      $&      0.053 &   40\\
     $H$ &  RRab &    18.913$\pm$0.064      &$     -2.214\pm0.308      $&      0.059 &   36\\
      &   RRc &    18.373$\pm$0.152      &$     -2.886\pm0.311      $&      0.045 &    5\\
      &   All &    18.820$\pm$0.025      &$     -2.655\pm0.106      $&      0.046 &   39\\
   $K_s$ &  RRab &    18.901$\pm$0.056      &$     -2.259\pm0.264      $&      0.052 &   35\\
    &   RRc &    18.437$\pm$0.143      &$     -2.695\pm0.309      $&      0.073 &    5\\
    &   All &    18.868$\pm$0.033      &$     -2.418\pm0.143      $&      0.053 &   40\\
\enddata
\tablecomments{The zero-point ($a$), slope ($b$), dispersion ($\sigma$) and the number of stars ($N$) in the final PL fits are tabulated. The RRab sample also includes RRd stars with their fundamentalized periods.}
\end{deluxetable}

\section{Distance to Carina Dwarf Spheroidal} \label{sec_mu0}

We derived the distance modulus, $\mu_0$, to Carina dSph using the mean $JHK_s$-band magnitudes presented in subsection \ref{sec_meanmag} for three different types of pulsating stars, namely RR Lyrae, ACep and DCep. To obtain distance modulus, we fit the calibrator PLRs or PLZRs separately to the F and FO pulsators, as well as combining them by fundamentalizing the overtone periods, where applicable. All derived $\mu_0$ are reported in Table \ref{tab_mu}, where the errors quoted are only weighted mean errors.

\subsection{RR Lyrae} \label{sec_rrmu}

To apply RR Lyrae PLZRs, the metallicity for a given stellar system, expressed as $[\mathrm{Fe/H}]$, needs to be known a priori. For Carina dSph, we adopted $[\mathrm{Fe/H}]=-2.13\pm0.28$~dex based on the work of \citet{coppola2015} and \citet{braga2022}. Furthermore, V182 is known to be a peculiar RR Lyrae \citep{coppola2013,coppola2015}, hence it is discarded in the subsequent $\mu_0$ determination. For this analysis, we used RRd stars with their FO mode periods because the calibrations provided in \citet{bhardwaj2023} are based on RRc+d sample. As can be seen from Table \ref{tab_mu}, $\mu_0$ using the $J$-band PLZRs are larger than their couterparts in the $HK_S$-band, regardless of the subsamples used (pulsation modes of ab only, cd only, or ab$+$cd). This behavior is also seen in the Draco dSph in Paper I. In addition, $\mu_0$ determined using the cd subsample are also larger than other two subsamples, but also exhibit larger errors due to lower statistics. These results remain unchanged even if we use F-mode periods for RRd stars. To investigate these and crosscheck, we replaced the empirical PLZRs adopted from \citet{bhardwaj2023} with the theoretical PLZRs available in \citet{marconi2015}. The resulted $\mu_0$, which were also reported in Table \ref{tab_mu}, also exhibit a similar behavior, suggesting these might be intrinsic to the RR Lyrae calibrations in the NIR. Using independent empirical and theoretical calibrations, we also noted that $\mu_0$ derived are in excellent agreement for various combinations of filters and pulsation modes, except for the ab and ab$+$cd subsamples in the $K_s$-band where the difference is of 0.05~mag. 

Based on the above findings, we determined the RR Lyrae-based distance modulus as $\mu_0^{RRL}=20.079\pm0.028$~mag (statistical error only) using the $HK_s$-band ab$+$cd PLZRs, as recommended in \citet{bhardwaj2023}.

\begin{deluxetable}{lccc}
  \tabletypesize{\scriptsize}
  \tablecaption{Extibction-corrected distance moduli obtained using NIR photometry for pulsating stars in Carina dSph.}
  \label{tab_mu}
  \tablewidth{0pt}
  \tablehead{
    \colhead{Type} &
    \colhead{$J$} &
    \colhead{$H$} &
    \colhead{$K_S$} 
  }
  \startdata
  \multicolumn{4}{c}{RR Lyrae, empirical PLZRs} \\
  ab      & $20.10\pm0.03$ & $20.05\pm0.03$  & $20.04\pm0.03$  \\
  cd      & $20.16\pm0.05$ & $20.14\pm0.05$  & $20.14\pm0.05$  \\
  ab$+$cd & $20.13\pm0.03$ & $20.07\pm0.03$  & $20.05\pm0.03$  \\
  \multicolumn{4}{c}{RR Lyrae, theoretical PLZRs} \\
  ab      & $20.10\pm0.03$ & $20.08\pm0.03$  & $20.09\pm0.03$  \\
  cd      & $20.17\pm0.05$ & $20.12\pm0.05$  & $20.12\pm0.05$  \\
  ab$+$cd & $20.12\pm0.02$ & $20.08\pm0.02$  & $20.10\pm0.02$  \\
  \hline
  \multicolumn{4}{c}{ACep, LMC calibration} \\
  F      & $19.81\pm0.08$ &  $\cdots$ & $19.86\pm0.07$ \\
  FO     & $19.87\pm0.07$ &  $\cdots$ & $19.87\pm0.06$ \\
  F+FO   & $19.80\pm0.06$ &  $\cdots$ & $19.84\pm0.05$ \\
  \multicolumn{4}{c}{ACep, SMC calibration} \\
  F      & $19.91\pm0.07$ &  $\cdots$ & $19.98\pm0.07$ \\
  FO     & $19.87\pm0.10$ &  $\cdots$ & $19.89\pm0.11$ \\
  F+FO   & $19.85\pm0.06$ &  $\cdots$ & $19.92\pm0.06$ \\
  \hline
  \multicolumn{4}{c}{DCep, SX Phe PLRs} \\
  F       & $20.15\pm0.05$  & $\cdots$  & $20.07\pm0.06$ \\
  \multicolumn{4}{c}{DCep, DSCT PLZRs, $[\mathrm{Fe/H}]=-1.7$~dex} \\
  F       & $20.08\pm0.04$  & $20.08\pm0.04$  & $20.08\pm0.04$ \\
  \multicolumn{4}{c}{DCep, DSCT PLZRs, $[\mathrm{Fe/H}]=-2.0$~dex} \\
  F       & $20.02\pm0.04$  & $20.04\pm0.04$  & $20.04\pm0.04$ \\
  \enddata  
\end{deluxetable}

\subsection{Anomalous Cepheids} \label{sec_acmu}

\begin{figure}
  \epsscale{1.1}
  \plotone{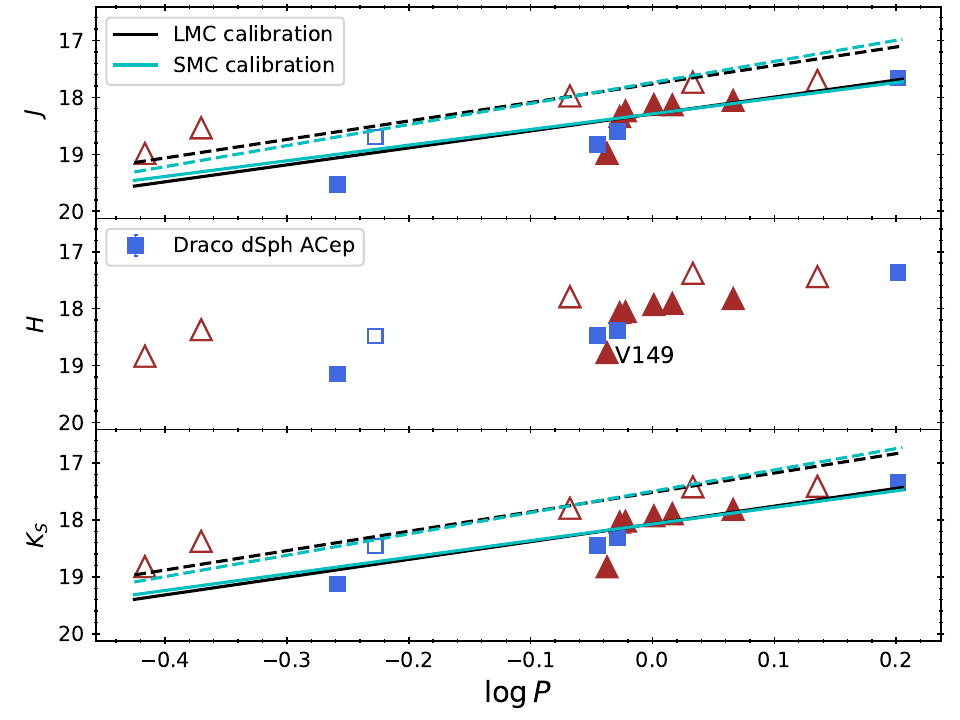}
  \caption{The extinction-corrected NIR PLRs for ACep, separated to fundamental (F) mode ACep and first-overtone (FO) mode ACep with filled and open symbols, respectively. The solid (for F ACep) and dashed (for FO ACep) lines represent the LMC (black lines) and SMC (cyan lines) PLRs adopted from \citet{sicignano2025}, after fitting for the respective distance moduli as given in Table \ref{tab_mu}. The blue squares represent ACep in Draco dSph from Paper I, shifted to the distance of Carina dSph.}
  \label{fig_acpl}
\end{figure}

Similar to RR Lyrae, ACep can pulsate either in the F or the FO mode. Since \citet{coppola2015} did not classify the ACep into different pulsation modes, we classified them as done in the Appendix, and summarized in Table \ref{tab_meanmag} for the ACep observed in our work.     

Recently, \citet{sicignano2025} derived a set of optical to NIR PLRs, including the $JK_s$-band, for both the F and the FO pulsators based on nearly 200 ACep found in the Magellanic Clouds. These PLRs were derived separately for the Large Magellanic Cloud (LMC) and the Small Magellanic Cloud (SMC) variables. We adopted the precise distances measured in \citet[][$\mu_{LMC}=18.477\pm0.026$~mag]{pietrzynski2019} and \citet[][$\mu_{SMC}=18.977\pm0.032$~mag]{graczyk2020}, respectively, to calibrate the LMC and SMC zero-points. When combining the samples of F and FO pulsators, the periods of FO pulsators were fundamentalized using $\log P_F = \log P_O + 0.145$ \citep{sicignano2025}. The fitted PLRs are shown in Figure \ref{fig_acpl}.

In the Figure \ref{fig_acpl}, an outlier ACep (V149) appears in the PLRs (which is $\sim 2.4\sigma$ away from the fitted PLRs). However, this ACep is not an anomaly in the CMD nor in the period-Wesenheit plane (see the right panel of Figure \ref{fig_acmode}), with respect to other ACep. Examining the right panel of Figure \ref{fig_acmode} revealed that the dispersions of the fundamental mode ACep in both of the Carina dSph and LMC are comparable. V149 and five other F ACep observed in this work are located at both ends of the period-Wesenheit plane, though the former has poorer light curve sampling as well.  If V149 is excluded, the resulting $\mu_0$ is systematically $\sim 1\sigma$ smaller.  

Compared to $\mu_0^{RRL}$ derived in the previous subsection, ACep-based $\mu_0$ tend to be smaller (from $\sim0.1$ to $\sim 0.3$~mag, see Table \ref{tab_mu}), which is also seen in Draco dSph. As discussed in Paper I and \citet{sicignano2025}, such a difference hints at a metallicity dependence on the PLRs for ACep. This is further supported by comparing the $\mu_0$ derived using the LMC vs. SMC PLRs, where the SMC-based $\mu_0$, for the F and F+FO subsamples, are closer to $\mu_0^{RRL}$. The difference in $\mu_0$ is a $\sim 0.11$~mag for the F pulsators when applying the LMC and SMC PLRs. 

Since there is a lack of metallicity measurement for ACep in the Magellanic Clouds \citep{sicignano2025}, we estimated the metallicity dependence on the ACep PLRs using two types of metallicity indicators or stellar populations in the Magellanic Clouds. Using red giant branch stars, metallicities for the LMC and SMC are found to be $-0.42$~dex \citep{choudhury2021} and $-0.97$~dex \citep{choudhury2020}, respectively. This translates to a metallicity dependence of $-0.20$~mag/dex in the ACep PLRs. On the other hand, \citet{muraveva2025} measured [Fe/H]$\sim-1.63$~dex and $-1.86$~dex using the old-population RR Lyrae in the LMC and SMC, respectively, and correspondingly the metallicity dependence on ACep PLRs is estimated to be $-0.48$~mag/dex. These two values suggest ACep PLRs are sensitive to metallicity as found in Paper I ($\sim -0.34$ mag/dex) and in \citet[][$\sim -0.26$ mag/dex]{sicignano2025}.

In either way, such metallicity dependence implies that ACep in Carina dSph might have a lower metallicity than ACep in the SMC. This is based on the hypothesis that a PLR calibrated from a stellar system with metallicity lower than SMC might bring the ACep-based $\mu_0$ in agreement with $\mu_0^{RRL}$. Considering the metallicity dependence of $-0.20$~mag/dex and $-0.48$~mag/dex, the metallicity for ACep in Carina dSph is estimated to be $\sim-1.64$~dex and $\sim-2.14$~dex, respectively. The recent spectroscopic investigation by \citet{ripepi2024} has revealed that ACep are generally metal-poor, with $\textrm{[Fe/H]}<-1.5$~dex.

Interestingly, $\mu_0$ using only the FO pulsators give very consistent values with $\mu_0\sim 19.87$~mag \citep[similar to the low value available in the literature, $19.87\pm0.11$~mag as found in][]{mighell1997}, independent of whether LMC or SMC PLRs were used as calibrators. This may suggest that the PLRs for FO pulsators are less sensitive to metallicity. However, their discrepancy with respect to RR Lyrae based distance suggests possible uncertainties in mode classification as well as effects of low statistics.

Finally, applying the theoretical $K$-band PLRs presented in \citet[][their equation 2]{fiorentino2006}, we obtained $\mu_0$ to be $20.00\pm0.06$~mag and $20.04\pm0.08$~mag for the F and FO pulsators, respectively. These distance moduli are in better agreement to the RR Lyrae-based $\mu_0$ than the empirical results. 

\subsection{Dwarf Cepheids} \label{sec_dcmu}

\begin{figure}
  \epsscale{1.1}
  \plotone{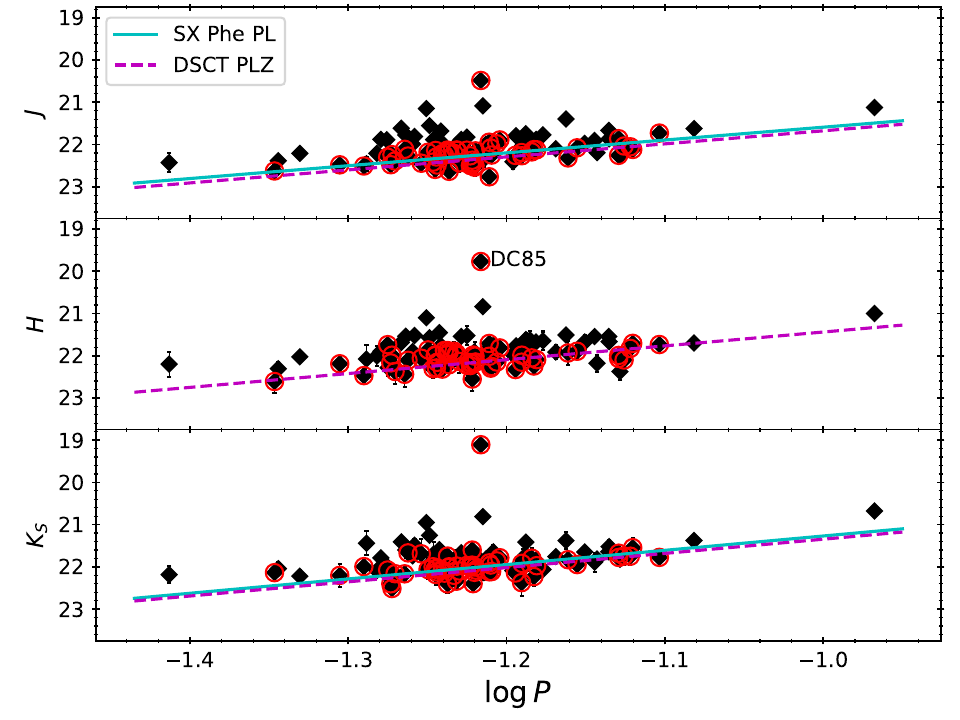}
  \caption{The extinction-corrected PLRs for $\sim 100$ DCep (black symbols) observed with the FourStar camera. Symbols with a red circle are the fundamental mode DCep identified in \citet{vivas2013}, which were also used to fit the SX Phe PLRs and the DSCT PLZRs at $[\mathrm{Fe/H}]=-1.7$~dex, after excluding the outlier DC85. The fitted $\mu_0$ in each filter are reported in Table \ref{tab_mu}.}
  \label{fig_dcpl1}
\end{figure}

Based on their locations on the period-luminosity planes, \citet{vivas2013} separated the DCep in Carina dSph into fundamental mode (F) and non-fundamental mode pulsators. Out of $\sim 100$ DCep detected in our FourStar observations, 56 are F pulsators. Except DC85, these fundamental mode DCep exhibit a clear sequence in the period-luminosity planes, as shown in Figure \ref{fig_dcpl1}. We adopted this subsample of the DCep in fitting the PLRs because, in general, the mode identification for DCep is not as straight forward as in the case of RR Lyrae \citep{vivas2013}. The outlier DC85 is located very close to a bright star, hence its photometry is severely blended, causing it to have unusual bright and red photometry (see Figure \ref{fig_cmd}). Therefore, we excluded this outlier in the following $\mu_0$ determination.    

\begin{deluxetable*}{lll}
  \tabletypesize{\scriptsize}
  \tablecaption{Summary of $\mu_0$ measurements to Carina dSph since 2015.}
  \label{tab_2015}
  \tablewidth{0pt}
  \tablehead{
    \colhead{$\mu_0$ [mag.]} &
    \colhead{Method} &
    \colhead{Reference} 
  }
  \startdata
  $19.98\pm0.03$ & $VI$-band RR Lyrae PW relation & \citet{Nagarajann2022} \\
  $20.02\pm0.05$ & $BV$-band RR Lyrae PW relation & \citet{coppola2015} \\
  $20.08\pm0.05$ & $gr$-band magnitude of RR Lyrae \& RC stars  & \citet{santana2016} \\
  $\sim20.11$    & ZAHB tracks on optical CMD & \citet{vdb2015} \\
  $20.12\pm0.11$ & $JK$-band RR Lyrae PLZ relation & \citet{karczmarek2015} \\
  \enddata
  \tablecomments{The abbreviations are: PW = Period-Wesenheit; RC = Red Clump; ZAHB = Zero-Age Horizontal-Branch.}
\end{deluxetable*}

As discussed in \citet{vivas2013}, DCep include both the population II, metal-poor SX Phe stars and the population I DSCT stars that have a higher metallicity. Dwarf galaxies such as Carina dSph could host both populations, which are difficult to separate. Hence, empirical PLRs or PLZRs for both of the SX Phe and DSCT available in the literature were used to fit our sample of fundamental mode DCep. Only SX Phe PLRs in the NIR were available from \citet{navarrete2017}, who derived a set of $JK_s$-band PLRs using 45 fundamental mode SX Phe stars in the globular cluster $\omega$ Centauri. As listed in Table \ref{tab_mu}, the derived $\mu_0$ from the SX Phe $JK_s$-band PLRs are consistent with each other within uncertainties. 

Recently, \citet{liu2025} published multiband PLRs and PLZRs for a sample of field DSCT, which have measured spectroscopic metalliticity. Their PLRs and PLZRs were calibrated using the Gaia DR3 parallaxes. Since DSCT stars in general have a higher metallicity than SX Phe stars, we adopted two metallicities, $[\mathrm{Fe/H}]=-1.7$~dex and $[\mathrm{Fe/H}]=-2.0$~dex, as in \citet{vivas2013} when fitting the $JHK_s$-band PLZRs. As can be seen from Table \ref{tab_mu}, the derived $\mu_0$ in the $JHK_s$ bands are remarkably in good agreement. We did not apply the PLRs derived in \citet{liu2025}, because the mean metallicity for their sample of DSCT is $-0.28$~dex. Applying the PLZRs at this mean metallicity gives $\mu_0\sim20.3$~mag.   

\begin{figure}
  \epsscale{1.1}
  \plotone{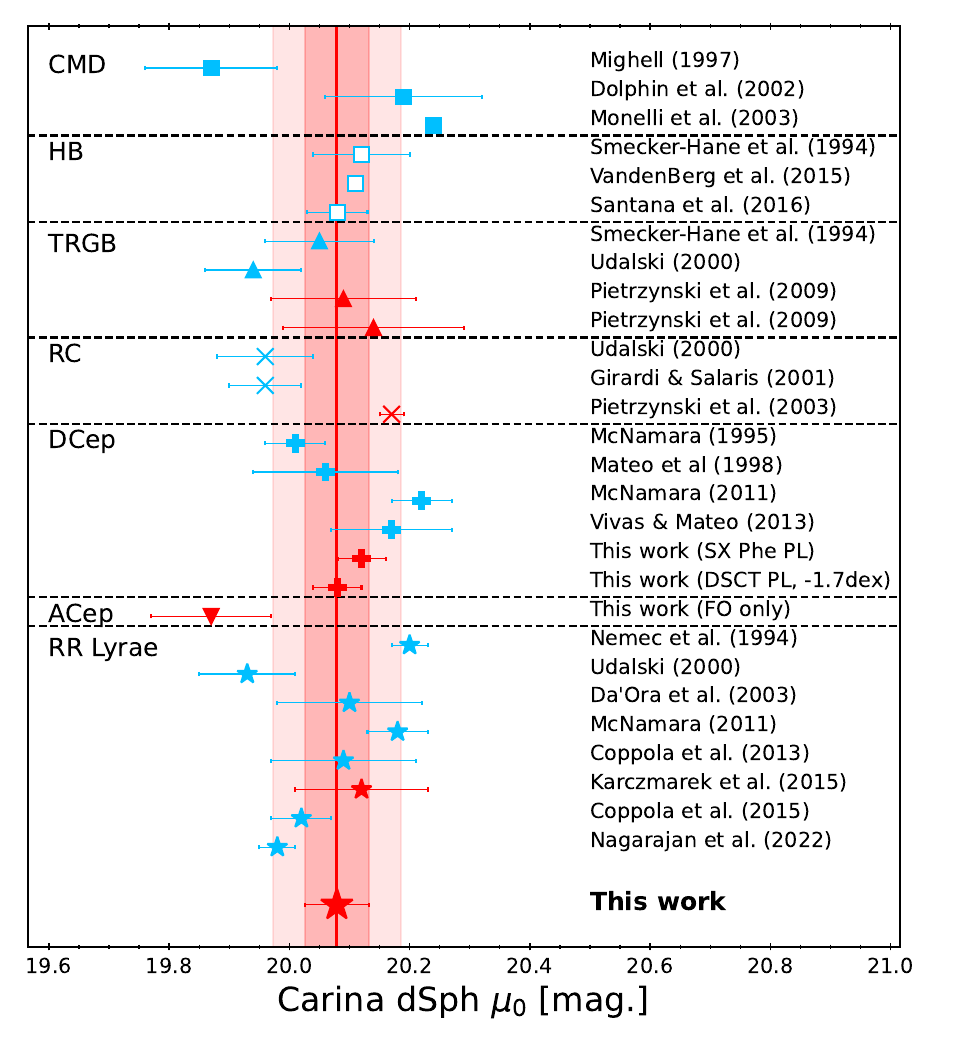}
  \caption{Comparison of our derived $\mu_0^{RRL}$ with the literature values as compiled in \citet{karczmarek2015} and in Table \ref{tab_2015}. The dark and light shaded areas represent  $\pm1\sigma$ and $\pm2\sigma$ (total error) of $\mu_0^{RRL}$, respectively. The blue and red colors represent the optical and the NIR measurements, respectively. Methods involved in the published measurements include: CMD \citep[filled squares,][]{mighell1997,dolphin2002,monelli2003}, horizonal-branch (HB) stars (including RR Lyrae) on the CMD \citep[open squares,][]{sh1994,vdb2015} or including red-clump (RC) stars \citep{santana2016}, tip of the red-giant branch \citep[TRGB; triangles,][]{sh1994,udalski2000,pietrzynski2009}, RC stars \citep[crosses,][]{udalski2000,girardi2001,pietrzynski2003}, DCep \citep[plus symbols,][]{mcnamara1995,mateo1998,mcnamara2011,vivas2013}, and RR Lyrae via PLRs, PLZRs, or period-Wesenheit relations \citep[star symbols,][]{nemec1994,udalski2000,dallora2003,mcnamara2011,coppola2013,karczmarek2015,coppola2015,Nagarajann2022}. For ACep, we adopted the $\mu_0$ based on FO pulsators, which gives consistent value of $\sim 19.87$~mag.}
  \label{fig_mucomp}
\end{figure}

\section{Discussion and Conclusion} \label{sec_end}

In this work, we presented NIR time-series observations of two Carina dSph fields using the FourStar camera mounted on the Magellan Telescope, with the main goal of obtaining high-precision distance to this dwarf galaxy using pulsating stars such as RR Lyrae. Our observations recovered 43, 11, and 102 of the known RR Lyrae, ACep and DCep, respectively.

We adopted the distance modulus determined using RR Lyrae (see subsection \ref{sec_rrmu}) as our final distance measurement to Carina dSph, which gives $\mu_0^{RRL}= 20.079\pm0.028\ \mathrm{(statistical)}\pm0.045\ \mathrm{(systematic)}$~mag. The systematic error includes the zero-point errors of the calibrator PLZRs ($\sim 0.02$~mag), the error on metallicity ($0.021$~mag) assuming an uncertainty of $0.1$~dex, as well as the zero-point errors on the empirical Carina dSph PLRs ($\sim0.033$~mag). Zero-point calibration uncertainties are included both in statistical and systematic error budgets. Therefore, final adopted distance modulus translates to a distance of $103.7\pm1.3\ \mathrm{(statistical)}\pm 2.2\ \mathrm{(systematic)}$~kpc, representing a distance error of $\sim 2.1\%$ (systematic) to Carina dSph.

Our final adopted $\mu_0^{RRL}$  is consistent with the earlier work of \citet{karczmarek2015}, albeit with significantly better precision, as this previous study used a smaller sample of 33 RRL and with photometry obtained from single-epoch $JK$-band observations. \citet{karczmarek2015} have also summarized the $\mu_0$ measurements in the literature prior to their work, spanning from 19.87~mag to 20.24~mag.\footnote{Note that \citet{karczmarek2015} did not include the earlier measurements from \citet[][$20.20\pm0.03$~mag]{nemec1994} and \citet[][$20.19\pm0.13$~mag]{dolphin2002}.} In Table \ref{tab_2015}, we have collected other measurements $\mu_0$ since 2015, including the work of \citet{karczmarek2015}. Our derived $\mu_0^{RRL}$ is broadly consistent with these modern measurements. Comparison of our derived $\mu_0^{RRL}$ with other literature values is shown in Figure \ref{fig_mucomp}. 

In case of ACep, we found that the derived $\mu_0$ are in general smaller than $\mu_0^{RRL}$, regardless of the filters, pulsation modes, or using the LMC/SMC calibrated PLRs. In the Paper I, a metallicity coefficient of $-0.3$ mag/dex was predicted for the ACep PLRs. In this work, the difference in distance moduli calibrated with LMC and SMC PLRs was quantified to be $-0.2$ or $-0.5$~mag/dex due to metallicity difference. However, such difference due to metallicity dependence is not clear for the first overtone pulsators, which also resulted in a smaller $\mu_0$ of $\sim19.87$~mag. The smaller statistics of ACep and possible contamination in mode identifications can {\bf result in} systematic differences in distance determinations. Because of these reasons, we did not adopt the distance measurements based on ACep.

Due to the ambiguous nature of DCep, we fitted the empirical PLRs or PLZRs derived from both of the SX Phe and DSCT populations to the DCep detected from our FourStar median-combined images. Nevertheless, the fitted $\mu_0$ are consistent with each other (especially in the $K_s$-band) and to $\mu_0^{RRL}$. Given the complex stars formation history of Carina dSph, with at least two episodes of star formation to be responsible for the old and intermediate age populations \citep[e.g., see][]{vdb2015,santana2016,norris2017}, it is entirely possible that DCep population in Carina dSph is a mix of the SX Phe and DSCT populations.

\begin{acknowledgments}
We thank the useful discussions and comments from an anonymous referee to improve the manuscript. We sincerely thank Yuri Beletsky who carried the remote Magellan obervation for our program. CCN thanks the funding from the National Science and Technology Council (NSTC, Taiwan) under the grant 114-2112-M-008-011. AB thanks the funding from the Anusandhan National Research Foundation (ANRF) under the Prime Minister Early Career Research Grant scheme (ANRF/ECRG/2024/000675/PMS). This research was supported by the International Space Science Institute (ISSI) in Bern/Beijing through ISSI/ISSI-BJ International Team project ID $\#$24-603 – ``EXPANDING Universe'' (EXploiting Precision AstroNomical Distance INdicators in the Gaia Universe).

  This research has made use of the SIMBAD database and the VizieR catalogue access tool, operated at CDS, Strasbourg, France. This research made use of Astropy,\footnote{\url{http://www.astropy.org}} a community-developed core Python package for Astronomy \citep{2013A&A...558A..33A,2018AJ....156..123A,2022ApJ...935..167A}. This paper includes data gathered with the 6.5 meter Magellan Telescopes located at Las Campanas Observatory, Chile. Our Magellan telescope time was granted by the Academia Sinica Institute of Astronomy and Astrophysics (ASIAA) and NSF NOIRLab (via NOIRLab Proposal ID 2025A-518184) through the Telescope System Instrumentation Program (TSIP). TSIP was funded by the U.S. National Science Foundation. 

This publication makes use of data products from the Two Micron All Sky Survey, which is a joint project of the University of Massachusetts and the Infrared Processing and Analysis Center/California Institute of Technology, funded by the National Aeronautics and Space Administration and the National Science Foundation. 

\end{acknowledgments}



%

\facilities{Magellan:Baade (FourStar)}

\software{{\tt FSRED},
  {\tt DAOPHOT/ALLSTAR} \citep{1987PASP...99..191S}, {\tt ALLFRAME} \citep{1994PASP..106..250S},
  {\tt dustmaps} \citep{green2018}
}

\appendix

\section*{Mode Identification for ACep in Carina dSph} \label{appxa}

\begin{figure*}
  \epsscale{1.1}
  \plottwo{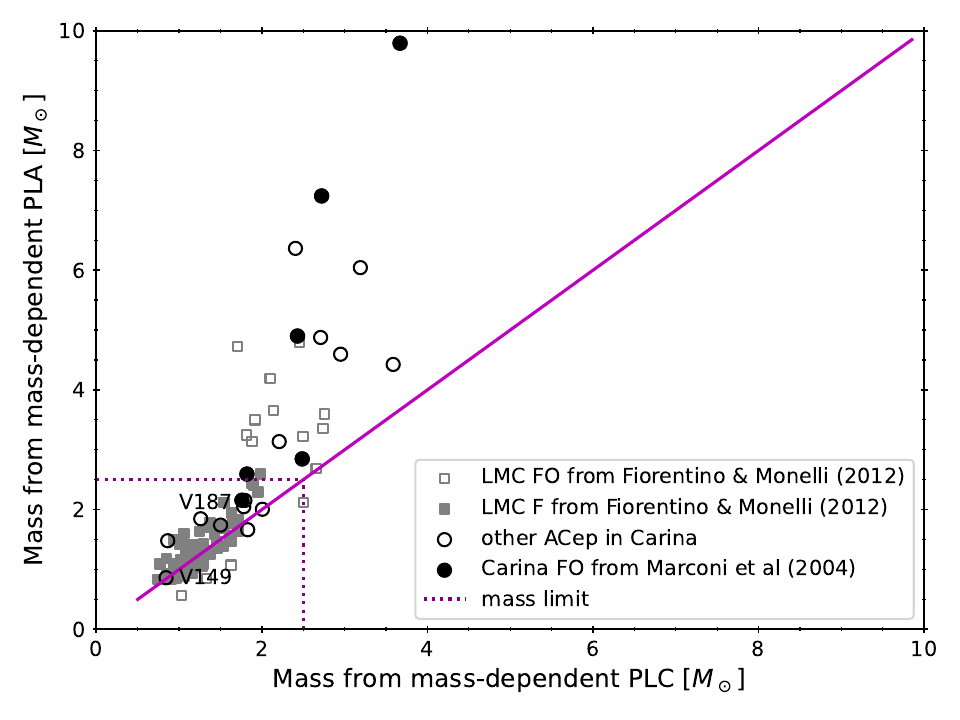}{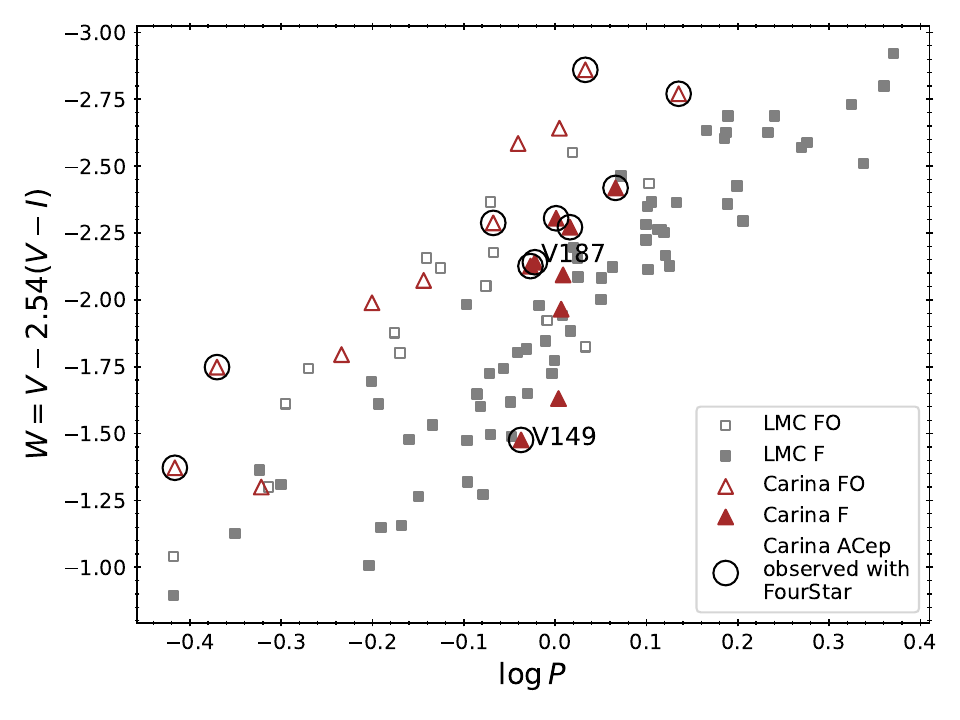}
  \caption{{\it Left Panel:} Mass estimations for ACep in Carina dSph (circles) using the theoretical mass-dependent PLC and PLA relations, and those classified in \citet{marconi2004} are marked with filled circles. The solid line represents the one-to-one relation, and the dotted lines are the adopted mass limit to select fundamental mode pulsators. Refined classification of LMC ACep \citep{fiorentino2012} are shown as grey squares for a comparison. {\it Right Panel:} The period-Wesenheit relation for ACep in both of the Carina dSph (triangles) and LMC (squares), with filled and open symbols representing the F and FO pulsators, respectively. ACep included in this work were further marked with an open circle. A distance modulus of  $20.079$~mag (see subsection \ref{sec_rrmu}) and $18.477$~mag \citep{pietrzynski2019} was adopted to convert the observed Wesenheit magnitudes to absolute magnitudes for the Carina dSph and LMC ACep, respectively.}
  \label{fig_acmode}
\end{figure*}

Based on a grid of ACep pulsation models, \citet{marconi2004} derived a set of mass-dependent period-luminosity-colors (PLC) relations for both of the F and FO ACep pulsators. These theoretical relations provide an opportunity to estimate the mass of a ACep once its distance and extinction is known. In addition, these authors have also derived a mass-dependent period-luminosity-amplitude (PLA) relation, which is only valid for the fundamental mode ACep. Hence, \citet{marconi2004} proposed the pulsation mode for ACep can be identified by comparing the masses derived from the mass-dependent PLC and PLA relations. That is, if the derived masses are consistent from using these two relations, then the ACep should be pulsating in fundamental mode, else it would be a first-overtone pulsator. Using such an approach, \citet{marconi2004} identified the pulsation modes for a number of ACep in local dwarf galaxies, including Carina dSph. This method was also adopted by \citet{fiorentino2012} to refine the pulsation modes for LMC ACep.   

By adopting the $VI$-band photometry and $V$-band amplitudes from \citet{coppola2015}, together with the RR Lyrae-based $\mu_0$ derived in subsection \ref{sec_rrmu}, we compared the masses of all of the ACep in Carina dSph in the left panel of Figure \ref{fig_acmode} (circles) by assuming all of them are F pulsators. Overplotted on this plot are the LMC ACep (grey squares) with pulsation modes  refined by \citet{fiorentino2012} for a comparison. As can be seen from this plot, masses estimated using either the mass-dependent PLC or PLA relations are consistent with each other for masses below $\sim 2.5M_\odot$. Furthermore, ACep have masses between $\sim 1M_\odot$ and $\sim 2.5M_\odot$. Therefore, we identified a given Carina ACep as a F pulsator if both of the masses determined using two approaches are smaller than $2.5M_\odot$ (the dotted lines shown in the left panel of Figure \ref{fig_acmode}), otherwise it would be a FO pulsator.    

\citet{marconi2004} identified two and six ACep in the Carina dSph as F and FO pulsators, respectively. However, the two F pulsators (V158 and V182) were classified as RR Lyrae in both \citet{dallora2003} and \citet{coppola2015}. Among the six FO pulsators, we recovered five of them. Based on the more extensive data available from \citet{coppola2015}, V187 is found to be pulsating in fundamental mode rather in the first overtone mode. As can be seen on the right panel of Figure \ref{fig_acmode}, there is no clear separation of F and FO ACep in the period-Wesenheit plane.


\bibliography{Carina_Magellan}{}
\bibliographystyle{aasjournal}



\end{document}